\documentclass{ws-procs975x65}

\begin{document}

\newcommand{\bra}[1]{\langle #1 |}
\newcommand{\ket}[1]{| #1 \rangle}
\newcommand{\braket}[2]{\langle #1 | #2\rangle}
\newcommand{\bbra}[1]{\bigl\langle #1 \bigr|}
\newcommand{\bket}[1]{\bigl| #1 \bigr\rangle}
\newcommand{\bbraket}[2]{\bigl\langle #1 \big| #2\bigr\rangle}
\newcommand{\Bbra}[1]{\Bigl\langle #1 \Bigr|}
\newcommand{\Bket}[1]{\Bigl| #1 \Bigr\rangle}
\newcommand{\Bbraket}[2]{\Bigl\langle #1 \Big| #2\Bigr\rangle}
\newcommand{\vev}[1]{\langle #1 \rangle}
\newcommand{\bvev}[1]{\bigl\langle #1 \bigr\rangle}
\newcommand{\Bvev}[1]{\Bigl\langle #1 \Bigr\rangle}
\newcommand{\dvev}[1]{\langle\!\langle #1 \rangle\!\rangle}
\newcommand{\bdvev}[1]{\bigl\langle\!\bigl\langle #1 \bigl\rangle\!\bigr\rangle}
\newcommand{\Bdvev}[1]{\Bigl\langle\!\Bigl\langle #1 \Bigr\rangle\!\Bigl\rangle}
\newcommand{\tvev}[1]{\langle\!\langle\!\langle #1 \rangle\!\rangle\!\rangle}
\newcommand{\btvev}[1]{\bigl\langle\!\bigl\langle\!\bigl\langle #1 \bigl\rangle\!\bigl\rangle\!\bigr\rangle}
\newcommand{\Btvev}[1]{\Bigl\langle\!\Bigl\langle\!\Bigl\langle #1 \Bigr\rangle\!\Bigr\rangle\!\Bigl\rangle}
\newcommand{\Ahat}{\hat{A}}
\newcommand{\Bhat}{\hat{B}}
\newcommand{\Chat}{\hat{C}}

\newcommand{\0}{\underline{0}}
\newcommand{\1}{\underline{1}}
\renewcommand{\i}{\underline{i}}
\renewcommand{\a}{\underline{a}}
\renewcommand{\b}{\underline{b}}
\renewcommand{\c}{\underline{c}}
\renewcommand{\d}{\underline{d}}
\newcommand{\g}{\underline{g}}
\renewcommand{\x}{\underline{x}}
\newcommand{\y}{\underline{y}}
\newcommand{\z}{\underline{z}}
\newcommand{\F}[1]{\mathbb{F}_{#1}}
\newcommand{\inner}[2]{\bigl(#1,#2\bigr)}
\newcommand{\dotproduct}[2]{#1\!\cdot\!#2}
\newcommand{\ppm}{\varphi}
\newcommand{\ppmf}[1]{\ppm\left(#1\right)}
\newcommand{\GFsigma}{\hat{\sigma}}

\title{Quantum Systems based upon Galois Fields\\
-- from Sub-quantum to Super-quantum Correlations --}

\author{L. N. Chang, Z. Lewis, D. Minic and T. Takeuchi\footnote{Presenting Author}}

\address{Department of Physics, Virginia Tech, Blacksburg, VA 24061, USA\\
E-mail: laynam@vt.edu, zlewis@vt.edu, dminic@vt.edu, takeuchi@vt.edu}

\begin{center}
{Dedicated to Freeman Dyson on the occasion of his 90th birthday}
\end{center}

\begin{abstract}
In this talk we describe our recent work on discrete quantum theory based on Galois fields. In particular, we discuss how
discrete quantum theory sheds new light on the foundations of quantum theory and we review
an explicit model of super-quantum correlations we have constructed in this context. We also discuss the 
larger questions of the origins and foundations of quantum theory, as well as the relevance of super-quantum theory
for the quantum theory of gravity.
\end{abstract}



\bodymatter

\section{Introduction: Why the quantum?}\label{aba:sec1}

Quantum theory is at present the most fundamental framework for physics.
Quantum theory underlies condensed matter, molecular, atomic, nuclear and particle physics,
as well as the cosmology of the early universe, spanning many orders of magnitude in scale.
However, the deep foundations of quantum theory are still an active area of investigation, even after more than
80 years\footnote{Counting from the 5th Solvay Conference in 1927.} since its inception.
In particular, it is not clear what the simplest logical underpinnings of quantum theory really are and how
unavoidable those assumptions might really be. 
As John Wheeler put it, we are still grappling with the question: ``Why the quantum?''

At the moment we have prescriptions for the ``quantization'' of a physical system and the ``interpretation'' of the resulting mathematics, but beyond that do we truly understand what quantum theory means?
The following famous quotations reminds us of the gravity of our situation:
``For those who are not shocked when they first come across quantum theory cannot possibly have understood it'' (attributed to Niels Bohr) or
``I think I can safely say that nobody understands quantum mechanics,'' as claimed by Richard Feynman.

Quantum theory predicts the probabilities of possible outcomes of a measurement. 
But what is ``measurement?''
Quantum theory does not provide a definition.
Also, what is ``probability?'' 
Is probability frequency? 
If so, how can one associate a probability to a physical event that happens only once?
Then is it Bayesian?
If so, whose subjective probability does it represent?
And how is quantum probability different from its classical counterpart?
Similarly, how does deterministic classical mechanics emerge from probabilistic quantum theory? (Decoherence? Many-worlds? Pilot waves?)
Finally, where should we draw the line between the observer and the observed? 
What if the observed is the entire Universe? Again, should we invoke many-worlds, or something else?
The questions are never-ending, and we can get pretty philosophical about it, exempli gratia:
\begin{quote}
``I cannot help thinking that our awareness of our own brains has something to do with the process which we call `observation' in atomic physics.  That is to say, I think our consciousness is not just a passive epiphenomenon carried along by chemical events in our brains, but is an active agent forcing the molecular complexes to make choices between one quantum state and another.  In other words, mind is already inherent in every electron, and the process of human consciousness differ only in degree but not in kind from the processes of choice between quantum states which we call `chance' when they are made by electrons.''
\end{quote}
as Freeman Dyson
wrote in his wonderful book ``Disturbing the Universe.''\cite{Dyson}

Such profound thoughts aside, it is far from clear whether the quantum formalism is adequate
for the understanding of many outstanding questions in physics concerning 
quantum gravity, 
the nature of the initial state of the Universe, 
the deep meaning of space and time, 
the origin of the Standard Model of matter, 
the nature of dark energy and dark matter, et cetera, et cetera.
To address these questions,
do we need to transcend the fundamental framework of quantum theory at some point, and if so, where, and how?
If not, why not, and what does that say about the foundations of quantum theory?
Also, where should we look for the necessary empirical evidence for such a new framework?
Finally, would such a new framework of physics shed new light on the deep structure of quantum theory as
well on its relation with its classical counterpart?

Note that neither quantum field theory (QFT), which is just a larger version of quantum theory, nor String Theory (as currently understood) challenge the basic tenets of quantum theory.
As David Gross reminded us in his talk at the 2005 Solvay conference: 
\begin{quote}
``
Many of us believed that string theory was a very dramatic break with our previous notions of quantum theory. But now we learn that string theory, well, is not that much of a break. The state of physics today is like it was when we were mystified by radioactivity. They were missing something absolutely fundamental. We are missing perhaps something as profound as they were back then.''
\end{quote}
Could it be that going beyond the quantum framework is the key?

What is the best way to address all these questions?
More phenomenology, that is, confronting the predictions of quantum theory with experiment, 
would probably not tell us anything more than what we already know: quantum theory works!
Instead, we propose that one should compare the predictions of quantum theory with those of its ``mutants,''
id est, theories whose mathematical structures have been slightly modified from 
the canonical version.
By looking at which physical predictions change under each ``mutation'' and which do not, one
can expect to bring to the surface the deeper connections between the mathematical structure
and the physical characteristics of the theory, and eventually provide answers to questions such as:
Is quantum theory inevitable?
Can quantum theory be derived from a few basic physical principles that everyone can agree on, \`a la Relativity?  
In particular, can the mathematical axioms of quantum theory be derived from those physical principles? 
And furthermore, by modifying those principles can one go beyond canonical quantum theory, opening the way to
the quantization of gravity? (And perhaps also explain dark energy, dark matter and the origin of the Standard Model of visible matter?)
Once we know how to go beyond quantum theory, we can 
then envision creating a new phenomenology in which post-quantum phenomena play a central role.

In the following, we review our recent work on ``mutant'' quantum theories constructed on
discrete and finite vector spaces over Galois fields.\cite{Chang:2012eh,Chang:2012gg,Chang:2012we,Takeuchi:2012mra} 
(See also Refs.~\citen{Nambu,Finkelstein,Penrose,MQT}.) 
Our models are necessarily ``mutant'' given that these vector spaces
do not possess inner products, and the formalism of canonical quantum theory cannot be applied as is.
Being discrete and finite the models are very simple, yet they turn out to be extremely illuminating.
We first set the stage in section~2 by a brief discussion of Bell's inequalities,\cite{bell} which 
accentuate the distinction between the classical and quantum worlds, 
and also serves to characterize a possible post-quantum world via super-quantum correlations.
In sections 3 and 4, we review our discrete ``mutant'' models with sub-quantum and super-quantum correlations, respectively.
Section 5 discusses what remains un-mutated in our models, while
section 6 summarizes the lessons learned from our toy models and points out an avenue for future work.
In the final section, we conclude by outlining the relevance of post-quantum theory for the foundations of quantum gravity\cite{Chang:2010ir,Chang:2013lsa}, and, in particular, string theory \cite{Chang:2011yt,Chang:2011jj,Freidel:2013zga}.

\section{Correlations in Classical and Quantum theories, and beyond}

Here, we briefly review the essence of 
Bell's seminal contributions to the foundation of quantum theory,
which can be viewed as, perhaps, the
simplest argument that distinguishes classical from quantum physics. 
In reviewing this classic argument, 
we also point out the logical possibility for theories beyond quantum theory,
i.e. theories characterized by correlations that are stronger than that of quantum theory, 
which we label ``super-quantum'' theories.

According to the celebrated Bell's inequalities,\cite{bell}
or its slightly generalized version, the Clauser-Horne-Shimony-Holt (CHSH) inequality
\cite{Clauser:1969ny} which we review here,
classical and quantum correlations are clearly separated by $O(1)$ effects. 
Let $A$ and $B$ represent the outcomes 
of measurements performed on some isolated physical system by detectors 
1 and 2 which are placed at two causally disconnected 
spacetime locations.  Assume that the only possible values of $A$ 
and $B$ are $\pm 1$.  Let $P(a,b)=\vev{A(a)B(b)}$ be the expectation value 
of the product $A(a)B(b)$ where $a$ and $b$ respectively denote 
the settings of detectors 1 and 2. Then, the upper 
bound, $X$, of the following combination of correlators, for arbitrary 
detector settings {$a$, $a'$, $b$, $b'$}, characterizes each underlying theory:
\begin{equation}
\Bigl|
 P(a, b) 
+P(a, b')
+P(a', b) 
-P(a', b') 
\Bigr| \;\leq\; X\;.
\label{CHSHbound}
\end{equation}
This bound for classical hidden variable theories is $X_\mathrm{Bell}=2$, while 
that for quantum theory is $X_\mathrm{QM}=2\sqrt{2}$.\cite{cirelson} 
That is, quantum mechanical correlations violate 
the classical Bell bound but are themselves bounded.  
Proofs are reviewed in Ref.~\citen{Chang:2011yt}.
From purely statistical reasoning one can conclude that the maximum 
possible value of $X$ is 4, and it has been demonstrated that the requirement of 
relativistic causality does not preclude correlations which saturate this absolute 
bound \cite{super}.

The question then arises whether there exist theories with super-quantum correlations, i.e. theories that violate
the quantum bound of $2 \sqrt{2}$. 
If such theories are not forbidden, then they must be compulsory, to cite
Murray Gell-Mann from a different context. 
Furthermore, in Refs.~\citen{Chang:2013lsa} and \citen{Chang:2011yt} 
we have argued that quantum gravity may necessarily be such a super-quantum theory.
This was the prime motivation for our search for a simple model with super-quantum correlations
in the context of discrete quantum theories over Galois fields \cite{Chang:2012eh,Chang:2012gg,Chang:2012we,Takeuchi:2012mra}.

Note that the CHSH inequality relies on the knowledge of expectation values and not probabilities.
Though predicting probabilities and predicting expectation values may seem like the same thing,
it turns out they are not necessarily when ``mutations'' are introduced.
To obtain a super-quantum theory, 
one must focus on the requirement that it is the predictions for the expectation values that should saturate the bound of 4. 
This is what we have done in Ref.~\citen{Chang:2012we}, which will be reviewed in section~4.
If one focusses on predicting probabilities, one obtains a different ``mutant'' 
which we will review first in the next section.
\cite{Chang:2012eh,Chang:2012gg,Takeuchi:2012mra}

\section{Galois Field Quantum Mechanics}


\subsection{The Mutation}

In canonical quantum theory, the states of an $N$-level quantum system are described by vectors
in the Hilbert space $\mathcal{H}_\mathbb{C}=\mathbb{C}^N$.\footnote{%
We restrict our attention to pure states.}
Here,
we introduce a ``mutation'' by replacing $\mathcal{H}_\mathbb{C}$ with 
$\mathcal{H}_q=\F{q}^N$,\footnote{%
A similar proposal was made by	
Schumacher and Westmoreland.\cite{MQT}
In their work, probabilities were not defined.
Our model would correspond to assigning equal probabilities to
all `possible effects' in their model.
}
where $\F{q}$ is shorthand for the finite Galois field $GF(q)$,
$q=p^n$ for some prime $p$, and $n\in\mathbb{N}$.
For the case $n=1$, we have $\F{p}=GF(p)=\mathbb{Z}/p\mathbb{Z}=\{\0,\1,\cdots,\underline{p-1}\}$.
Such replacements of the vector space have been considered previously, 
\textit{e.g.} real quantum mechanics in which $\mathcal{H}_\mathbb{C}$ is replaced by $\mathcal{H}_\mathbb{R}=\mathbb{R}^N$,\cite{Stueckelberg:1960}
and quaternionic quantum mechanics in which it is replaced by $\mathcal{H}_\mathbb{H}=\mathbb{H}^N$.\cite{Adler:1995}
However,
the vector space $\mathcal{H}_q$, in contradistinction to $\mathcal{H}_\mathbb{R}$, $\mathcal{H}_\mathbb{C}$,
or $\mathcal{H}_\mathbb{H}$, 
lacks an inner product, normalizable states, and symmetric/hermitian operators.
Nevertheless, we find that we can construct a quantum-like model on it,
which predicts probabilities of physical measurements that cannot be
reproduced in any hidden variable theory.
What will not survive this ``mutation,'' however, are the
correlations of canonical quantum theory that violate
the classical CHSH bound of $X_\mathrm{Bell}=2$.

\subsection{The Model}

As discussed at the end of the previous section, we would like to construct a model
which predicts probabilities for the outcomes of measurements.
Our starting point is the following canonical expression for the
probability of obtaining the outcome represented by the dual-vector $\bra{x}\in\mathcal{H}_\mathbb{C}^*$
when a measurement is performed on the state represented by the 
vector $\ket{\psi}\in\mathcal{H}_\mathbb{C}$: 
\begin{equation}
P(x|\psi) \;=\; \dfrac{\bigl|\braket{x}{\psi}\bigr|^2}{\sum_y \bigl|\braket{y}{\psi}\bigr|^2}\;.
\label{Pdef}
\end{equation}
Here, $\ket{\psi}$ is not normalized and
the sum in the denominator runs over the duals of all the eigenstates of a hermitian operator
which represents the observable in question.
However, for this expression to be interpretable as a probability,
the necessary condition is that the dual-vectors in the sum
span the entire dual vector space $\mathcal{H}_\mathbb{C}^*$,
and any reference to operators acting on $\mathcal{H}_\mathbb{C}$ is inessential.
The interpretation that the bracket $\braket{x}{\psi}\in\mathbb{C}$ is
an inner product between two vectors also need not be imposed.
The probability depends only on the absolute values of the brackets
$|\braket{x}{\psi}|\in\mathbb{R}$.
Since we can multiply $\ket{\psi}$ with any non-zero complex number
without changing the probabilities defined via Eq.~(\ref{Pdef}),
we are compelled to identify
vectors which differ by a non-zero multiplicative constant
as representing the same physical state, endowing the
state space with the complex projective geometry
\begin{equation}
\mathbb{C}P^{N-1}
\;=\; (\,\mathbb{C}^{N}\backslash\{\mathbf{0}\}\,)\,\big/\,(\,\mathbb{C}\backslash\{0\}\,)
\;\cong\; S^{2N-1}\!\big/\,S^1\;,
\label{CPN}
\end{equation}
where each line going through the origin of $\mathbb{C}^N$ is identified as a `point.'

Thus, to construct a ``mutant"' quantum theory on $\mathcal{H}_q$, 
we represent states with vectors $\ket{\psi}\in\mathcal{H}_q$, 
and outcomes of measurements with dual-vectors $\bra{x}\in\mathcal{H}_q^*$.
Observables are associated with a choice of basis of $\mathcal{H}_q^*$,
each dual-vector in it representing a different outcome.
The bracket $\braket{x}{\psi}\in\F{q}$ is converted into
a non-negative real number $|\braket{x}{\psi}|\in\mathbb{R}$ 
via the absolute value function:
\begin{equation}
|\,\underline{k}\,|\;=\;
\left\{\begin{array}{ll}
0\quad &\mbox{if $\underline{k}=\0$}\;,\\
1\quad &\mbox{if $\underline{k}\neq\0$}\;.
\end{array}
\right.
\label{abs}
\end{equation}
Here, underlined numbers and symbols represent elements of $\F{q}$, to
distinguish them from elements of $\mathbb{R}$ or $\mathbb{C}$.
Note that Eq.~(\ref{abs}) is not to be interpreted as a condition imposed on $\braket{x}{\psi}\in\F{q}$;
all non-zero values of $\F{q}$ are mapped to one.
Since $\F{q}\backslash\{\0\}$ is a cyclic multiplicative group, 
this assignment of `absolute values' is the only one consistent with 
the requirement that the map from $\F{q}$ to non-negative $\mathbb{R}$ be product preserving,
that is: $|\underline{k}\underline{l}|=|\underline{k}||\underline{l}|$.\footnote{%
The product preserving nature of the absolute value function 
guarantees that the probabilities of product observables on product states
factorize in multi-particle systems.
This property is crucial if we want to have isolated particle states,
and is of course shared by canonical quantum theory defined on $\mathcal{H}_{\mathbb{C}}$.
}
With these assignments, Eq.~(\ref{Pdef}) can be applied as it stands to calculate probabilities.
Since the same absolute value is assigned to all non-zero brackets,
all outcomes $\bra{x}$ for which the bracket with the state $\ket{\psi}$
is non-zero are given equal probabilistic weight.

Note also that the multiplication of $\ket{\psi}$ with a non-zero element of
$\F{q}$ will not affect the probability. 
Thus,
vectors that differ by non-zero multiplicative constants are identified as representing the same 
physical state, and the state space is endowed with the finite projective geometry
\cite{Hirschfeld,Hirschfeld2,Arnold}
\begin{equation}
PG(N-1,q) \;=\; (\,\F{q}^N\backslash\{\mathbf{\0}\}\,)\,\big/\,(\,\F{q}\backslash\{\0\}\,)\;,
\label{PGdef}
\end{equation}
where each `line' going through the origin of $\F{q}^N$ is identified as a `point,'
in close analogy to the complex projective geometry of canonical QM.

\subsection{An Example}

To give a concrete example of our proposal, let us construct a 2-level system,
 analogous to spin, for which $\mathcal{H}_q = \F{q}^2$, and the state space is
$PG(1,q)$.
This geometry consists of $q+1$ `points,'
which can be represented by the vectors
\begin{equation}
\ket{\,0\,} = \left[\begin{array}{c} \1 \\ \0 \end{array}\right],\quad
\ket{\,1\,} = \left[\begin{array}{c} \0 \\ \1 \end{array}\right],\quad
\ket{\,r\,} = \left[\begin{array}{l} \g^{r-1} \\ \1 \end{array}\right],
\end{equation}
$r=2,3,\cdots,q$, where $\g$ is the generator of the multiplicative group $\F{q}\backslash\{\0\}$
with $\g^{q-1}=1$.
The number $q+1$ results from the fact that of the $q^2-1$ non-zero vectors, every $q-1$ are equivalent,
thus the number of inequivalent vectors are $(q^2-1)/(q-1)=(q+1)$.
Similarly, the $q+1$ inequivalent dual-vectors can be represented as:
\begin{eqnarray}
\bra{\,\overline{0}\,} & = & \bigl[\,\0\;-\!\1\,\bigr]\;,\cr
\bra{\,\overline{1}\,} & = & \bigl[\,\1\;\;\phantom{-}\0\,\bigr]\;,\cr
\bra{\,\overline{r}\,} & = & \bigl[\,\1\;-\!\g^{r-1}\,\bigr]\;,\qquad r=2,3,\cdots,q\;,
\end{eqnarray}
where the minus signs are dropped when the characteristic of $\F{q}$ is two.
From these definitions, we find:
\begin{eqnarray}
\braket{\bar{r}}{s} 
& =    & \0\quad \mbox{if $r=s$}\;, \cr 
& \neq & \0\quad \mbox{if $r\neq s$}\;,
\end{eqnarray}
and
\begin{equation}
\bigl|\braket{\bar{r}}{s}\bigr|\;=\; 1-\delta_{rs}\;.
\label{braketpq}
\end{equation}
Observables are associated with a choice of basis of $\mathcal{H}_q^*$:
\begin{equation}
A_{rs}\;\equiv\;\{\;\bra{\bar{r}},\;\bra{\bar{s}}\;\}\;,\qquad
r\neq s\;.
\end{equation}
We assign the outcome $+1$ to the first dual-vector of the pair, 
and the outcome $-1$ to the second to make these observables
spin-like. This assignment implies $A_{sr}=-A_{rs}$.
The indices $rs$ can be considered as indicating the direction of the `spin,'
and the interchange of the indices as indicating a reversal of this direction.
Mappings of these `spin' directions to actual directions in 3D space are
discussed in Ref.~\citen{Chang:2012gg}.

Applying Eq.~(\ref{Pdef}) to this system, it is straightforward to show that
\begin{eqnarray}
P(A_{rs}=+1\,|\,r) & = & 0\;,\qquad
P(A_{rs}=-1\,|\,r) \;=\; 1\;,\cr
P(A_{rs}=+1\,|\,s) & = & 1\;,\qquad
P(A_{rs}=-1\,|\,s) \;=\; 0\;,\cr
P(A_{rs}=\pm 1\,|\,t) & = & \frac{1}{2}\;,\qquad\mbox{for $t\neq r, s$}\;,
\end{eqnarray}
and thus,
\begin{eqnarray}
\vev{A_{rs}}_r & = & -1\;,\cr
\vev{A_{rs}}_s & = & +1\;,\cr
\vev{A_{rs}}_t & = & \phantom{-}0\;,\quad\mbox{for $t\neq r, s$.}
\end{eqnarray}
So for each `spin,' there exist two `eigenstates,'
one for $+1$ (`spin' up) and another for $-1$ (`spin' down).
For all other states the two outcomes $\pm 1$ are equally probable.

The states and observables `rotate' into each other under changes of bases.
For the projective geometry $PG(1,q)$, the group of all possible basis transformations
constitute the projective group $PGL(2,q)$ of order $q(q^2-1)$.
$PGL(2,q)$ is formally a subgroup of $S_{q+1}$, the group of all possible permutations of the $q+1$ states.

\subsection{Spin Correlations}

To show that our system is ``quantum'' in the sense that no hidden variable theory can reproduce
its predictions, we use an argument 
analogous to those of Greenberger, Horne, Shimony, and Zeilinger,\cite{GHZ}
and of Hardy\cite{Hardy:1993zza} for canonical quantum theory.
Let us construct
a two `spin' system on the tensor product space $\F{q}^2\otimes\F{q}^2 = \F{q}^4$.
The number of non-zero vectors in this space is
$q^4-1$, of which every $q-1$ are equivalent, so the number of inequivalent
states is $(q^4-1)/(q-1)=q^3+q^2+q+1$.
Of these, $(q+1)^2$ are product states, leaving
$(q^3+q^2+q+1)-(q+1)^2=q(q^2-1)$ that are entangled.
As noted in footnote d, Eq.~(\ref{Pdef}) applied to tensored spaces
with the product preserving absolute value function Eq.~(\ref{abs}) 
ensures that the expectation values of product observables
factorize for product states, thereby rendering the distinction between
product and entangled states meaningful.

The number of entangled states matches the order of the group $PGL(2,q)$,
since arranging the 4 elements of an entangled state into a $2\times 2$ array
gives rise to a non-singular matrix.
The entangled states fall into `conjugacy' classes, matching those of $PGL(2,q)$,
that transform among themselves under $PGL(2,q)$ `rotations.'
The singlet state, corresponding to the conjugacy class of the unit element,
can be expressed as
\begin{equation}
\ket{S} \;=\; \ket{r}\otimes\ket{s}-\ket{s}\otimes\ket{r}\;,\qquad r\neq s\;,
\end{equation}
for any two states $\ket{r}$ and $\ket{s}$ up to a multiplicative constant.
If the characteristic of $\F{q}$ is two, the minus sign is replaced by a plus sign.

Products of the `spin' observables are defined as
\begin{equation}
A_{rs}A_{tu}
\,=\,\{
\,\bra{\bar{r}}\otimes\bra{\bar{t}}\,,
\,\bra{\bar{r}}\otimes\bra{\bar{u}}\,,
\,\bra{\bar{s}}\otimes\bra{\bar{t}}\,,
\,\bra{\bar{s}}\otimes\bra{\bar{u}}\,
\}\;,
\end{equation}
the four tensor products representing the outcomes
$++$, $+-$, $-+$, and $--$,
and the expectation value giving the correlation between the two `spins.' 
The probabilities of the four outcomes are
particularly easy to calculate for the singlet state $\ket{S}$ since \cite{MQT}
\begin{eqnarray}
\bigl(\bra{\bar{r}}\otimes\bra{\bar{s}}\,\bigr)\ket{S}
& =    & \0\quad \mbox{if $r=s$}\;, \cr
& \neq & \0\quad \mbox{if $r\neq s$}\;,
\end{eqnarray}
thus
\begin{equation}
\Bigl|
\bigl(\bra{\bar{r}}\otimes\bra{\bar{s}}\,\bigr)\ket{S}
\Bigr|
\;=\; 1-\delta_{rs}\;,
\end{equation}
and we obtain the probabilities and correlations listed in Table.~\ref{Probs}.

\begin{table}[t]
\begin{center}
\tbl{Probabilities and expectation values of product observables in the singlet state $\ket{S}$.
The indices $r$, $s$, $t$, and $u$ are distinct.
Cases that can be obtained by flipping signs using $A_{rs}=-A_{sr}$ are not shown.}
{\begin{tabular}{|c||c|c|c|c||c|}
\hline
\ Observable\ \ &\ $++$\ \ &\ $+-$\ \ &\ $-+$\ \ &\ $--$\ \ &\ E.V. \ \\
\hline
$A_{rs}A_{rs}$ & $0$            & $\dfrac{1}{2}$ & $\dfrac{1}{2}$ & $0$            & $-1$ \phantom{\bigg|} \\
\hline
$A_{rs}A_{rt}$ & $0$            & $\dfrac{1}{3}$ & $\dfrac{1}{3}$ & $\dfrac{1}{3}$ & $-\dfrac{1}{3}$ \phantom{\bigg|}\\
\hline
$A_{rs}A_{st}$ & $\dfrac{1}{3}$ & $\dfrac{1}{3}$ & $0$            & $\dfrac{1}{3}$ & $+\dfrac{1}{3}$ \phantom{\bigg|}\\
\hline
$A_{rs}A_{tu}$ & $\dfrac{1}{4}$ & $\dfrac{1}{4}$ & $\dfrac{1}{4}$ & $\dfrac{1}{4}$ & $\phantom{-}0$  \phantom{\bigg|}\\
\hline
\end{tabular}}
\label{Probs}
\end{center}
\end{table}

To demonstrate that these correlations cannot be reproduced in 
any hidden variable theory, it suffices to look at the correlations between two observables that share an index.
For instance, consider the following two:
\begin{equation}
X\;\equiv\;A_{01}\;,\quad
Y\;\equiv\;A_{02}\;.
\end{equation}
First, from the first row of Table~\ref{Probs} we can discern that 
\begin{equation}
\begin{array}{lll}
P(X_1X_2;++|S) & =\; P(X_1X_2;--|S) & =\; 0\;,\\
P(Y_1Y_2;++|S) & =\; P(Y_1Y_2;--|S) & =\; 0\;,
\end{array}
\end{equation}
where we have added subscripts to distinguish between the two `spins.'
This tells us that the pairs $(X_1X_2)$
and $(Y_1Y_2)$ are completely anti-correlated. 
Next, from the second row of Table~\ref{Probs}, we conclude:
\begin{equation}
P(X_1Y_2;++|S) \;=\; P(Y_1X_2;++|S) \;=\; 0\;,
\end{equation}
which means that if either one of the pairs $(X_1Y_2)$ and $(Y_1X_2)$ is
$+1$, then its partner must be $-1$.
Thus, the implications of either $X_1=+1$ or $X_1=-1$ would be:
\begin{equation}
\begin{array}{llll}
X_1=+1 &\rightarrow\;
Y_2=-1 &\rightarrow\;
Y_1=+1 &\rightarrow\;
X_2=-1 \;,\\
X_1=-1 &\rightarrow\;
X_2=+1 &\rightarrow\;
Y_1=-1 &\rightarrow\;
Y_2=+1 \;.\\
\end{array}
\end{equation}
In either case, we cannot classically have $(X_1Y_2)=(--)$ 
or $(Y_1X_2)=(--)$, even though both configurations have quantum mechanical
probabilities of $1/3$.
Thus, the predictions of our ``mutant'' model do not
allow any hidden variable mimic.

To calculate the CHSH bound for our model, it suffices to examine all possible correlators
for the singlet state $\ket{S}$ only.
This is because all $q(q^2-1)$ entangled states can be transformed
into $\ket{S}$ via local $PGL(2,q)$ rotations, that is, $PGL(2,q)$ transformations on only one of 
the entangled particles.
Using the numbers listed in Table.~\ref{Probs}, it is then not difficult to
convince oneself that the CHSH bound for this model is the `classical' 2.\cite{Chang:2012eh,Chang:2012gg}

\subsection{Classical Limit?}

The model discussed in this section serves as an existence proof that quantum-like
theories whose predictions cannot be reproduced by any classical hidden variable theory
can nevertheless have correlations that are sub-quantum and do not violate the classical CHSH bound of
$X_\mathrm{Bell}=2$.
Thus, the absence of hidden variable mimics does not guarantee the violation of the classical CHSH bound.

We have yet to unravel any deep reason for this, but we have made one curious observation:
If we take the limit $q\rightarrow 1$, the model reduces to that defined on a
`vector space' over $\F{1}$, `the field with one element.' \cite{Tits:1957,Kurokawa:2010}
There, the projective geometry of the state space is preserved, but the superposition of states is forbidden.
The model becomes `classical' in the sense that only the eigenstates of only one observable
survive, the probability of any measurement yielding a particular result becoming either 0 or 1.
Perhaps it is not surprising then that the model for the $q\neq 1$ cases has the CHSH bound of 2, given
that it is independent of $q$, and the model reduces to a `classical' theory in the $q\rightarrow 1$ limit.
This observation also shows that $\hbar\rightarrow 0$ may not be the only path to
reach the `classical' limit of quantum-like theories.
Indeed, our model does not even have $\hbar$ in it.
Detailed discussions on these points will be presented in a separate publication.\cite{Fun}

\section{Biorthogonal Quantum Mechanics}

\subsection{Biorthogonal Systems}

The model presented in the previous section made use of Eq.~(\ref{Pdef}) to make 
contact with canonical quantum theory.
An alternative is to go through the canonical expression
\begin{equation}
\dfrac{\bra{\psi}\Ahat\ket{\psi}}{\braket{\psi}{\psi}}
\label{Avevdef}
\end{equation}
for the expectation value of the observable $\hat{A}$ on the state $\ket{\psi}$.
In canonical quantum theory on $\mathcal{H}_{\mathbb{C}}=\mathbb{C}^N$, 
$\bra{\psi}$ is the conjugate dual of the state $\ket{\psi}$ such that
\begin{equation}
\bra{\psi}\;=\;\inner{\ket{\psi}}{\phantom{\quad}}\;,
\end{equation}
where $\inner{\quad}{\quad}$ is the inner product of $\mathcal{H}_{\mathbb{C}}$, 
while $\hat{A}$ is required to be hermitian, that is:
\begin{equation}
\hat{A} \;=\; \sum_{k=1}^N \alpha_k\ket{k}\bra{k}\;,\qquad \alpha_k\in\mathbb{R}\;,
\end{equation}
for some orthonormal basis $\{\ket{1},\ket{2},\cdots,\ket{N}\}$ of $\mathcal{H}_{\mathbb{C}}$.
To make use of Eq.~(\ref{Avevdef}) in a model defined
on the vector space $\mathcal{H}_q=\F{q}^N$, which does not have an inner product, the `conjugate dual'
$\bra{\psi}$ of a state $\ket{\psi}$ and the analog of the hermitian operator $\hat{A}$ must
be defined judiciously.

For this purpose, we first restrict the Galois field over which the vector space is constructed
to the case $q=p^2$ with $p = 3\,\mathrm{mod}\,4$.
The Galois field $\F{p^2}$ is obtained from $\F{p}$ by adjoining the solution to $\underline{x}^2+\1=\0$
which we will denote $\i$.\footnote{%
If $p=2$ or $1\,\mathrm{mod}\,4$, then $\underline{x}^2+\1=\0$ is reducible, $\underline{p-1}$ providing a solution.} 
That is $\F{p^2}=\F{p}[\i]$.  
For example, if we write the elements of $\F{3}$ as $\F{3}=\{\1,\0,-\1\}$, then
$\F{9}=\F{3}[\i]=\{\1,\0,-\1,\i,-\i,\1+\i,\1-\i,-\1+\i,-\1-\i\}$.
Thus the pair $\F{p}$ and $\F{p^2}=\F{p}[\i]$ provides an
analog of the pair $\mathbb{R}$ and $\mathbb{C}=\mathbb{R}[i]$.

We next define the `dot product' in $\F{p^2}^N$ as
\begin{equation}
\dotproduct{\ket{a}}{\ket{b}}\;=\;
\sum_{k=1}^{N}\,
\a_k^p\,
\b_k^{\phantom{p}}
\;\in\;\F{p^2}\;,
\end{equation}
where $\a_k$ and $\b_k$ are respectively the $k$-th element of $\ket{a}$ and $\ket{b}$.
Raising an element to the $p$-th power is semilinear in $\F{p^n}$ for any $n\in\mathbb{N}$
since
\begin{equation}
(\a+\b)^p \;=\; (\a^p+\b^p)
\end{equation}
in a field of characteristic $p$.
When $n=1$, it is an identity transformation due to
Fermat's little theorem
\begin{equation}
a^{p-1}\;=\;1\;\mathrm{mod}\;p\;,\quad\forall a\in\mathbb{Z}\;.
\end{equation}
For the case $n=2$, $p=3\;\mathrm{mod}\,4$, it is an analogue of
complex conjugation in $\mathbb{C}$ since
the elements of $\F{p^2}=\F{p}[\i]$ can be expressed
as $\a+\i\,\b$, where $\a,\b\in\F{p}$,
and
\begin{equation}
(\a+\i\,\b)^p
\,=\, \a^p + \i^{p}\b^p
\,=\, \a-\i\,\b\;.
\end{equation}
(Note that $p$ is odd so that $\i^p=-\i$.)
Furthermore,
\begin{eqnarray}
(\a+\i\,\b)^p(\c+\i\,\d)
& = & (\a\c+\b\d)+\i(\a\d-\b\c)\;,\cr
(\c+\i\,\d)^p(\a+\i\,\b)
& = & (\a\c+\b\d)-\i(\a\d-\b\c)\;,
\end{eqnarray}
in particular,
\begin{equation}
(\a+\i\,\b)^p(\a+\i\,\b)
\;=\; \a^2+\b^2
\;\in\; \F{p}\;.
\end{equation}
Therefore, 
$\dotproduct{\ket{a}}{\ket{b}}$ and
$\dotproduct{\ket{b}}{\ket{a}}$ are `complex conjugates' 
of each other, while
$\dotproduct{\ket{a}}{\ket{a}}$
is `real.'
Borrowing from standard terminology,
we will say that two vectors in $\F{p^2}^N$ are `orthogonal' to each other
when they have a zero dot product,
and that a vector is `self-orthogonal' when it is orthogonal to itself.

Using this dot-product, we define
the `conjugate dual' vector of a non-self-orthogoal vector $\ket{\psi}$ 
as
\begin{equation}
\bra{\psi} \;\equiv\; 
\dfrac{\ket{\psi}\,\cdot}{\dotproduct{\ket{\psi}}{\ket{\psi}}}
\label{Vconjugation}
\end{equation}
where it is crucial that $\dotproduct{\ket{\psi}}{\ket{\psi}}\neq\0$
for $\bra{\psi}$ to exist.
Therefore, not all vectors in our vector space have conjugate duals.

To define the analogue of hermitian operators, we invoke the notion of
biorthogonal systems.\footnote{%
Biorthogonal systems have been discussed in Ref.~\citen{BQS}
in the context of PT Symmetric Quantum Mechanics \cite{PT}.
}
A biorthogonal system of $\F{p^2}^N$ is a set consisting of a basis 
$\{\ket{1},\ket{2},\cdots,\ket{N}\}$ of the vector space $\F{p^2}^N$,
and a basis $\{\bra{1},\bra{2},\cdots,\bra{N}\}$ of the dual vector space $\F{p^2}^{N*}$ 
such that
\begin{equation}
\braket{r}{s} 
\;=\; \underline{\delta}_{rs}
\;=\;
\begin{cases}
\;\0\quad & \mbox{if $r\neq s$}\;, \\
\;\1\quad & \mbox{if $r=s$}\;.
\end{cases}
\label{biortho}
\end{equation}
Such a system can be constructed by first
choosing a basis $\{\ket{1},\ket{2},\cdots,\ket{N}\}$ for $\F{p^2}^N$  
such that:
\begin{equation}
\dotproduct{\ket{r}}{\ket{s}}\;
\begin{cases}
\;\neq\; \0\qquad & \mbox{if $r=s$}\;, \\
\;=\;    \0\qquad & \mbox{if $r\neq s$}\;,
\end{cases}
\label{OrthoBasis}
\end{equation}
that is, all the basis vectors are orthogonal to each other,
but none are self-orthogonal.
Let us call such a basis an `ortho-nondegenerate' basis.
The simplest example of an ortho-nondegenerate basis would be such
that the $r$-th element of the $s$-th vector is given by
$\underline{\delta}_{rs}$,
proving that such a basis always exists.
On the other hand, not all bases satisfy this condition 
since $\F{p^2}^N$ typically has multiple 
self-orthogonal vectors other than the zero vector as alluded to above.
For each vector $\ket{r}$ in this basis, define its conjugate dual $\bra{r}$ via Eq.~(\ref{Vconjugation}).
Then, the set of dual vectors $\{\bra{1},\bra{2},\cdots,\bra{N}\}$
provides a basis for the dual vector space $\F{p^2}^{N*}$ which satisfies Eq.~(\ref{biortho}).

Given a biorthogonal system for $\F{p^2}^N$, we define the analog of a hermitian operator by
\begin{equation}
\hat{A} \;\equiv\; \sum_{k=1}^N \underline{\alpha}_k \ket{k}\bra{k}\;,\qquad
\underline{\alpha}_k\in\F{p}\;.
\label{Adef}
\end{equation}
Note that the `eigenvalues' $\underline{\alpha}_k$ of $\hat{A}$ are chosen in $\F{p}$, the analog of $\mathbb{R}$, 
not in $\F{p^2}$, the analog of $\mathbb{C}$.
Aside from the choice of these `eigenvalues,' one such operator can be defined for each biorthogonal system.
With this definition of $\hat{A}$, we can calculate the expression 
$\bra{\psi}\hat{A}\ket{\psi}\in\F{p}$ for any state $\ket{\psi}\in\F{p^2}^N$ for which a dual $\bra{\psi}\in\F{p^2}^{N*}$ exists.

To associate $\bra{\psi}\hat{A}\ket{\psi}\in\F{p}$ with the expectation value of a physical observable such as spin, 
we must map this quantity to $\mathbb{R}$.
We demand that this mapping from $\F{p}$ to $\mathbb{R}$ be product preserving, which is required for
eigenvectors of $\hat{A}$ to correspond to states with zero uncertainty,
and for the expectation value of product states to factorize in multi-particle systems.
Aside from the absolute value function discussed in section~3, there is one other such map when $p=3\,\mathrm{mod}\,4$.
This mapping can be constructed as follows.
First, denote the generator
of the multiplicative group $\F{p}\backslash\{\0\}$ by $\g$
and express the non-zero elements of $\F{p}$ as 
$\{\g,\g^2,\g^3,\cdots,\g^{p-1}=\1\}$.
Define:
\begin{equation}
\ppmf{\x}
\;=\;
\begin{cases}
\phantom{-}0\quad & \mbox{if $\x=\0$}\;, \\
+1\quad & \mbox{if $\x=\g^\mathrm{even}$}\;, \\
-1\quad & \mbox{if $\x=\g^\mathrm{odd}$}\;.
\end{cases}
\label{phidef}
\end{equation}
It is straightforward to show that $\ppmf{\a\b}=\ppmf{\a}\ppmf{\b}$.

To summarize, in this new ``mutation'' on $\mathcal{H}_{p^2}=\F{p^2}^N$,
observables $\hat{A}$ are defined for each biorthogonal system via Eq.~(\ref{Adef}).
We restrict ``physical'' states $\ket{\psi}$ to those for which the conjugate dual $\bra{\psi}$ can be defined,
which is actually equivalent to demanding that it belong to some biorthogonal system.
Then, the expectation value of the observable $\hat{A}$ when 
a measurement is performed on $\ket{\psi}$ is given by
\begin{equation}
\ppmf{\bra{\psi}\hat{A}\ket{\psi}}\;\in\;\mathbb{R}\;.
\end{equation}
Note that if $\ket{\psi}$ is multiplied by a non-zero constant in $\F{p^2}$, 
$\bra{\psi}$ will be multiplied by its inverse, so $\hat{A}$ and
$\bra{\psi}\hat{A}\ket{\psi}$ are not affected.  That is, 
states that differ by a multiplicative non-zero constant are all equivalent
as in the case of the model discussed in section 3.

\subsection{An Example}

Consider the vector space $\F{9}^2$.
There are $9^2-1=80$ non-zero vectors in this space, of which every $9-1=8$ are
equivalent.  So there are $80/8=10$ inequivalent states which can be taken to be:
\begin{equation}
\begin{array}{lllll}
\ket{a}=\left[\begin{array}{c} \1 \\ \0 \end{array}\right],
&\ket{c}=\left[\begin{array}{c} \1 \\ \1 \end{array}\right],
&\ket{e}=\left[\begin{array}{c} \1 \\ \i \end{array}\right],
&\ket{g}=\left[\begin{array}{c} \1 \\ \1+\i \end{array}\right],
&\ket{i}=\left[\begin{array}{c} \1 \\ \1-\i \end{array}\right],
\\
\ket{b}=\left[\begin{array}{c} \0 \\ \1 \end{array}\right],
&\ket{d}=\left[\begin{array}{r} \1 \\ -\1 \end{array}\right],
&\ket{f}=\left[\begin{array}{c} \1 \\ -\i \end{array}\right],
&\ket{h}=\left[\begin{array}{c} \1 \\ -\1-\i \end{array}\right],
&\ket{j}=\left[\begin{array}{c} \1 \\ -\1+\i \end{array}\right].
\end{array}
\label{F9kets}
\end{equation}
Of these, $\ket{a}$, $\ket{b}$, $\ket{c}$, $\ket{d}$, $\ket{e}$, and $\ket{f}$
have conjugate duals which are given by
\begin{equation}
\begin{array}{llllll}
\bra{a} & =\;\left[\begin{array}{cc} \1 & \;\0 \end{array}\right],\;
\quad &
\bra{c} & =\;\left[\begin{array}{cc} -\1 & -\1 \end{array}\right],\; 
\quad &
\bra{e}\;=\;\left[\begin{array}{cc} -\1 & \phantom{-}\i \end{array}\right],
\\
\bra{b} & =\;\left[\begin{array}{cc} \0 & \;\1 \end{array}\right],\;
\quad &
\bra{d} & =\;\left[\begin{array}{cc} -\1 & \phantom{-}\1 \end{array}\right],\;
\quad &
\bra{f}\;=\;\left[\begin{array}{cc} -\1 & -\i \end{array}\right],
\end{array}
\label{F9bras}
\end{equation}
while $\ket{g}$, $\ket{h}$, $\ket{i}$, and $\ket{j}$ do not and are therefore ``unphysical.''
The biorthogonal systems of this vector space are
\begin{equation}
\bigl\{\{\bra{a},\bra{b}\},\,\{\ket{a},\ket{b}\}\bigr\}
\;,\;\;
\bigl\{\{\bra{c},\bra{d}\},\,\{\ket{c},\ket{d}\}\bigr\}
\;,\;\;\mbox{and}\;\;
\bigl\{\{\bra{e},\bra{f}\},\,\{\ket{e},\ket{f}\}\bigr\}
\;.
\label{F9bio}
\end{equation}
From these, we can construct three spin-like observables with eigenvalues $\pm\1$:
\begin{equation}
\begin{array}{llll}
\1\;\ket{a}\bra{a} \hspace{-2px}& -\1\;\ket{b}\bra{b} 
& = \;
\left[\begin{array}{rr} \1 & \0 \\ \0 & -\1 \end{array}\right]
& \equiv \;\GFsigma_3\;,
\\
\1\;\ket{c}\bra{c} \hspace{-2px}& -\1\;\ket{d}\bra{d} 
& = \;
\left[\begin{array}{rr} \0 & \1 \\ \1 & \phantom{-}\0 \end{array}\right]
& \equiv \;\GFsigma_1\;,
\\
\1\;\ket{e}\bra{e} \hspace{-2px}& -\1\;\ket{f}\bra{f} 
& = \;
\left[\begin{array}{rr} \0 & -\i \\ \i & \0 \end{array}\right]
& \equiv \; 
\GFsigma_2
\;.
\end{array}
\label{F9obs}
\end{equation}
These are just the Pauli matrices with elements in $\F{9}$ instead of $\mathbb{C}$.
Then, the expectation values of $\GFsigma_1$, for instance, for the six physical states
will be given by
\begin{eqnarray}
\ppmf{\bra{a}\GFsigma_1\ket{a}} & = & \phantom{-}0\;,\cr
\ppmf{\bra{b}\GFsigma_1\ket{b}} & = & \phantom{-}0\;,\cr
\ppmf{\bra{c}\GFsigma_1\ket{c}} & = & \phantom{-}1\;,\cr
\ppmf{\bra{d}\GFsigma_1\ket{d}} & = & -1\;,\cr
\ppmf{\bra{e}\GFsigma_1\ket{e}} & = & \phantom{-}0\;,\cr
\ppmf{\bra{f}\GFsigma_1\ket{f}} & = & \phantom{-}0\;.
\end{eqnarray}
%

\subsection{Spin Correlations}

In order to look at the correlations of two `spins,' we 
construct a two particle system on the tensor product space
$\F{9}^2\otimes\F{9}^2=\F{9}^4$.
Of the $9^4-1=6560$ non-zero vectors in this space, every $9-1=8$ are
equivalent, so the number of inequivalent states is
$6560/8=820$.  Of these, $10^2=100$ are product states while
$820-100=720$ are entangled.  Of the entangled states, it turns out that 
504 are physical while 216 are unphysical.
See Ref.~\citen{Chang:2012we} for details.

The product spin operators are given by the Kronecker products of the
Pauli matrices we derived above.  For instance
\begin{equation}
\begin{array}{llll}
\GFsigma_1\otimes\GFsigma_1
& = 
\left[
\begin{array}{rrrr}
\0 & \0 & \0 & \phantom{-}\1 \\
\0 & \0 & \phantom{-}\1 & \0 \\
\0 & \phantom{-}\1 & \0 & \0 \\
\1 & \0 & \0 & \0
\end{array}
\right],
&
\qquad
\GFsigma_1\otimes\GFsigma_3
& = 
\left[
\begin{array}{rrrr}
\0 &  \0 & \phantom{-}\1 &  \0 \\
\0 &  \0 & \0 & -\1 \\
\1 &  \0 & \0 &  \0 \\
\0 & -\1 & \0 &  \0
\end{array}
\right], 
\\
\GFsigma_3\otimes\GFsigma_1
& =  
\left[
\begin{array}{rrrr}
\0 &  \phantom{-}\1 &  \0 &  \0 \\
\1 &  \0 &  \0 &  \0 \\
\0 &  \0 &  \0 & -\1 \\
\0 &  \0 & -\1 &  \0
\end{array}
\right], 
&
\qquad
\GFsigma_3\otimes\GFsigma_3
& =  
\left[
\begin{array}{rrrr}
\1 &  \0 &  \0 &  \0 \\
\0 & -\1 &  \0 &  \0 \\
\0 &  \0 & -\1 &  \0 \\
\0 &  \0 &  \0 &  \phantom{-}\1
\end{array}
\right].
\end{array}
\end{equation}
The CHSH bound for this model turns out to be the super-quantum 4.
To see this, it suffices to calculate the correlations for one physical state
which saturates this bound.
We take this to be
\begin{equation}
\ket{U}\;=\; \left[\begin{array}{c} \1 \\ \0 \\ \1 \\ \1+\i \end{array}\right]\;,\qquad
\bra{U}\;=\; \Bigl[\;\1\;\;\0\;\;\1\;\;\1-\i\;\Bigr]\;.
\end{equation}
It is straightforward to show that
\begin{eqnarray}
\bra{U}\GFsigma_1\otimes\GFsigma_1\ket{U}
& = & \bra{U}\GFsigma_1\otimes\GFsigma_3\ket{U} 
\;=\; \bra{U}\GFsigma_3\otimes\GFsigma_3\ket{U}
\;=\; -\1\;,\cr
\bra{U}\GFsigma_3\otimes\GFsigma_1\ket{U}
& = & \1\;,
\label{Ucorr}
\end{eqnarray}
and consequently,
\begin{eqnarray}
& & \Bigl|\;\ppmf{\bra{U}\GFsigma_1\otimes\GFsigma_3\ket{U}}
+\ppmf{\bra{U}\GFsigma_1\otimes\GFsigma_1\ket{U}}
\cr
& & \;\;
+\ppmf{\bra{U}\GFsigma_3\otimes\GFsigma_3\ket{U}}
-\ppmf{\bra{U}\GFsigma_3\otimes\GFsigma_1\ket{U}}
\;\Bigr|
\;=\; 4\;.
\end{eqnarray}
%

\subsection{Indeterminate Probabilities}

Note that in this model, the expectation values are predicted but the
probabilities are not.  For instance, 
from $\ppmf{\bra{U}\GFsigma_3\otimes\GFsigma_3\ket{U}}=-1$ we can conclude
that the probabilities that the measurement of $\GFsigma_3$ on both `spins'
would yield $++$, $+-$, $-+$, and $--$, respectively, must satisfy the relations
\begin{equation}
\begin{array}{l}
P(++|U)+P(+-|U)+P(-+|U)+P(--|U) = \phantom{-}1\;,\\
P(++|U)-P(+-|U)-P(-+|U)+P(--|U) = -1\;,
\end{array} 
\end{equation}
which imply
\begin{eqnarray}
0 & = & P(++|U) \;,\cr
0 & = & P(--|U) \;,\cr
1 & = & P(+-|U)+P(-+|U) \;,
\end{eqnarray}
but the model does not specify what $P(+-|U)$ and $P(-+|U)$ are separately.
While this may seem like a problem at first sight,
it is no more peculiar than canonical quantum theory itself which 
only predicts probabilities of outcomes, and not the results of
individual measurements.
This model only takes the indeterminacy of the theory one step further 
and does not predict the probabilities of individual outcomes but only 
the final expectation value.  Physically, this could correspond to a situation
in which the frequencies of the individual outcomes fluctuate and never
settles into definite probabilities, but the outcomes nevertheless conspire
to yield a well defined expectation value upon repeated measurements.

It is tempting to contemplate that the
general structure of biorthogonal systems  
and the indeterminate nature of probabilities is valid for more
general constructions of super-quantum theories, including the ones that we
expect to be relevant in the quantum theory of gravity.
We will have more to say about this later.

\section{Un-mutated Aspects}

Before we continue, let us comment on several aspects of canonical quantum theory
that are not ``mutated'' in the ``mutations'' discussed above.
This is to give us a perspective on how close our mutants are to the canonical,
yet possess distinguishing features.

\subsection{Probabilities and Expectation Values}

The point of contact between the mutant of section~3 and canonical quantum theory was
Eq.~(\ref{Pdef}), and that for the mutant of section~4 was Eq.~(\ref{Avevdef}).
Though Eqs.~(\ref{Pdef}) and (\ref{Avevdef}) are equivalent in canonical quantum theory,
we have seen that they are not in our mutants due to the lack of an inner product,
and the necessity of introducing a map from $\F{q}$ to $\mathbb{R}$ at some point to
make contact with experiment.

While it is theoretically possible to contemplate a departure from both Eqs.~(\ref{Pdef}) and (\ref{Avevdef}),
we choose to maintain one or the other in the mutations discussed above.
The reasons are multiple.
In addition to our desire to simply keep things under control, the fact that
probabilities and expectation values are given by quadratic forms of the wave-function
in canonical quantum theory 
can be supported via the generic nature of the Fisher metric on the space of measured events\cite{wootters}.
Experiments also support the robustness of the Born rule\cite{triple}.
Thus, in our initial probe into the world of mutant theories, 
it seems prudent to keep this aspect of canonical quantum theory intact.

Maintaining Eq.~(\ref{Avevdef}), as was done in section~4, also allows us to maintain
contact with QFT where all physical quantities are expressed in terms of
correlation functions.
In conformal QFT in particular, the formulation is from a purely algebraic viewpoint
and does not involve the use of Lagrangians, Hamiltonians, or Feynman rules.
Everything is defined in terms of correlation functions, and the familiar derivation of
the S-matrix in other QFT's involving the convolution of correlation functions with the wave-functions of
external probes is not even a well-defined concept.

\subsection{Projective Linear and Unitary Groups}

The two mutations we have been discussing in the previous sections both have the property that
state vectors which differ by a non-zero multiplicative constant in $\F{q}$ 
all represent the same physical state.
Thus the state space possesses the projective geometry $PG(N-1,q)$, as defined
in Eq.~(\ref{PGdef}), in close analogy to the $\mathbb{C}P^{N-1}$ geometry, Eq.~(\ref{CPN}), of
canonical quantum theory.
The one difference is that in the model of section~3 all states were physical 
but in the model of section~4 some were not.
This difference leads to a difference in the possible analogs of `unitary' transformations
in the two models.
In the model of section~3, the group of non-trivial basis transformations was the projective linear group
\begin{equation}
PGL(N,q) \;=\; GL(N,q)/Z(N,q)\;,
\end{equation}
where $GL(N,q)$ is the group of non-singular $N\times N$ matrices with elements in $\F{q}$,
while $Z(N,q)$ is its center consisting of the unit matrix multiplied by non-zero constants in $\F{q}$.
This is in direct analogy with canonical quantum theory where the group
of non-trivial basis transformations on $\mathbb{C}P^{N-1}$ was
\begin{equation}
SU(N) \;=\; U(N)/U(1)\;.
\end{equation}
In the model of section~4, however, basis transformations must be from
one biorthogonal system to another so not all transformations in $PGL(N,q)$ are allowed.
In the case of the $\F{9}^2$ model discussed above, the allowed transformations 
are given by $2\times 2$ matrices $\underline{U}$ with elements in $\F{9}$
which satisfy the condition
\begin{equation}
\underline{U}^\dagger \underline{U} \;=\; \pm \mathbf{\1}_{2\times 2}\;,
\end{equation}
with matrices that differ by a non-zero multiplicative constant identified.\footnote{%
See the appendix of Ref.~\citen{Chang:2012we} for details.}
These matrices constitute the projective unitary group $PU(2,9)$, which is a
subgroup of $PGL(2,9)$.
Though the group is different,
we can again see the close analogy with $SU(2)$ of canonical quantum theory.

\section{Summary and Comments}

Our work, reviewed in sections 3 and 4,
shows that it is possible to construct quantum-like theories on a vector space without an inner product, normalizable states, or symmetric/Hermitian operators in more than one way.
The probabilities predicted by our first mutant discussed in section~3 
could not be reproduced in any hidden variable theory. 
Nevertheless, the CHSH bound of the mutant was the ``classical'' 2.

The CHSH bound of our second mutant discussed in section~4 was the super-quantum 4. 
That model, though constructed on a discrete and finite vector space
in which not all states were `physical,' nevertheless provides an existence proof
that super-quantum theories can and do exist.
The crucial ingredient in the setup was the adoption of predicting the expectation values
instead of probabilities as the objective of the theory.
This led to definite expectation values but indefinite probabilities.

We note that super-quantum correlations have been discussed extensively in the literature (see Refs.~\citen{super} and \citen{trivial}).
There, attention has often been focused on the pathologies that may result from super-quantum correlations,
and the argument has been that perhaps nature rejects their existence to avoid such complications.
Our work is complementary to these efforts in that it provides a toy model which actually
predicts super-quantum correlations on which various ideas about such super-quantum theories can perhaps be tested.

Our model, which is based on expectation values instead of probabilities, also provides a contrast
to efforts in the foundations of quantum theory community, 
which attempt to construct canonical quantum theory from ground up based
solely on the concept of probability (see for example Ref.~\citen{fuchs}). 
We argue that even though canonical quantum theory may
be based solely on the concept of probability, super-quantum theory
does not have to be. 
This is reinforced by our experience in modern QFT
(especially the conformal QFT's) in which one operates solely with correlation functions
as alluded to above.

The two pathways to a quantum-like theory presented above differed
partly due to the necessity of introducing a map from $\F{q}$ to $\mathbb{R}$
at some point to make contact with physical reality.  
Application of the two constructions to Banach spaces \cite{banach}
would be a natural place to further clarify the difference between the
two approaches, do away with the product preserving map from $\F{q}$ to $\mathbb{R}$,
and search for models which may serve as closer representations of reality
where various quantum gravitational ideas to be discussed below can be explored.

\section{Quantum Gravity as a Super-Quantum Theory}

\subsection{Expectation Values over Probabilities}

Our work on discrete quantum theory over Galois fields presents perhaps the simplest model for
super-quantum correlations. 
Super-quantum correlations are realized in
the model together with a signature feature:
the physics of the model is entirely determined in terms of expectation values,
whereas the probabilities are, in general, indeterminate.
This feature is actually quite
natural, and desirable, from various point of view suggested by different 
approaches to quantum gravity.

We first recall our observation that theories based on expectation values 
meshes well with conformal field theories (CFT's).
As is well known, CFT's can be dual to quantum gravitational theories in certain
backgrounds, namely the AdS spaces \cite{adscft}, and also in the context of the
observed cosmological de Sitter spacetimes \cite{dscft}.

Furthermore, different approaches to non-perturbative quantum gravity and quantum cosmology \cite{gh,Hardy:2005fq,chia} suggest that the individual probability for specific measurements should be 
indeterminate, and that the observables in that context are different from
the usual observables found in canonical quantum theory.
The model considered here should be viewed as a concrete realization of this
general expectation.

Another exciting possibility that is being explored recently is that quantum gravity demands 
energy-momentum space to be dynamical\cite{Chang:2010ir,Chang:2011jj,Freidel:2013zga}.
This would have profound implications on the conceptual foundations of quantum gravity as well as on its phenomenology.\cite{Chang:2011yt,Chang:2013lsa,Chang:2010ir,Chang:2011jj,Freidel:2013zga} 
Dynamical energy-momentum space taken together with dynamical spacetime would demand 
a dynamical phase space and thus, quite naturally, dynamical Hilbert spaces and dynamical probabilities, as also expected on other grounds.\cite{chia}
That is, quantum probabilities themselves should change dynamically with the dynamics of the phase space,
implying indeterminate probabilities in quantum gravity theories.

Thus, our simple super-quantum model sheds new light on the search for the simplest set of reasonable axioms that
lead to canonical quantum theory, and the generalizations of which would tell us how to quantize gravity\cite{chia,Hardy:2001jk}.

\subsection{Double Quantization}

Further insight can be obtained from our discrete toy model. 
Specifically, given the fact that
in our toy model of super-quantum theory the probabilities of individual 
outcomes were indeterminate while the expectation values of observables were 
determinable, this suggests that super-quantum correlations would result from 
a theory in which probability distributions themselves are probabilistically determined, 
pointing to a ``double'' quantization. 
In particular, as we have conjectured in Ref.~\citen{Chang:2011yt}, 
since the process of quantization increases the CHSH bound from the classical $2$
to the quantum $2\sqrt{2}$,  
another step of ``quantization'' could further increase the bound 
by a factor of $\sqrt{2}$ to the super-quantum $4$.

What procedure would 
such a ``double'' quantization entail?  Quantization demands that correlation functions 
of operators be calculated via the path integral
\begin{equation}
\bvev{\,\Ahat(a)\Bhat(b)\,}  \;=\; \int \!Dx\;A(a,x)\,B(b,x)\,\exp\left[\dfrac{i}{\hbar} S(x)\right]
\;\equiv\;
A(a)\star B(b)
\;,
\label{AstarB}
\end{equation}
where $x$ 
collectively denotes the classical dynamical variables of the system. In 
a similar fashion, we can envision performing another step of 
quantization by integrating over ``paths'' of quantum operators to define 
correlators of ``super'' quantum operators
\begin{equation}
\bdvev{\,\hat\Ahat(a)\hat\Bhat(b)\,}  
\;=\; \int \!D\hat{\phi}\;\Ahat(a,\hat{\phi})\,\Bhat(b,\hat{\phi})\,\exp\left[\dfrac{i}{\tilde{\hbar}} \tilde{S}(\hat{\phi})\right]\;,
\end{equation}
where $\hat{\phi}$ collectively denotes the 
dynamical quantum operators of the system. Here, $\dvev{\hat\Ahat(a)\hat\Bhat(b)}$ is an 
operator.  To further reduce it to a number, we must 
calculate its expectation value in the usual way
\begin{equation}
\bdvev{\,\hat\Ahat(a)\hat\Bhat(b)\,}  
\quad\rightarrow\quad
\btvev{\,\hat\Ahat(a)\hat\Bhat(b)\,}  
\;=\;
\left\langle
\int \!D\hat{\phi}\;\Ahat(a,\hat{\phi})\,\Bhat(b,\hat{\phi})\,\exp\left[\dfrac{i}{\tilde{\hbar}} \tilde{S}(\hat{\phi})\right]
\right\rangle
\;,
\end{equation}
which would 
amount to replacing all the products of operators on the 
right-hand side with their first-quantized expectation values, or 
equivalently, replacing the operators with `classical' variables except with their 
products defined via Eq.~(\ref{AstarB}). Note that this is precisely 
the formalism of Witten's open string field theory (OSFT) \cite{witten}, in 
which the action for the `classical' open string field $\Phi$
is given formally as
\begin{equation}
S_W (\Phi) \;=\; \int \Phi \star Q_\mathrm{BRST} \Phi + \Phi \star \Phi \star \Phi\;,
\end{equation}
where $Q_\mathrm{BRST}$ is the open string 
theory BRST cohomology operator ($Q_\mathrm{BRST}^2=0$), and the star product is 
defined via a world-sheet path integral weighted with the Polyakov 
action and deformation parameter $\alpha'=\ell_s^2$.
The fully quantum OSFT is 
then, in principle, defined by yet another path integral in 
the infinite dimensional space of the open string field $\Phi$, 
i.e. 
\begin{equation}
\int D \Phi\,\exp\left[\dfrac{i}{g_s}S_W(\Phi)\right]\;,
\end{equation} 
where $g_s$ is the string coupling and all products 
are defined via the star-product. For reasons of unitarity, 
OSFT must contain closed strings, and therefore gravity. Thus, OSFT 
is a manifestly ``doubly'' quantized theory, and we argue that 
it, and the theory of quantum gravity it should become, 
would be characterized by super-quantum correlations when fully formulated.

In a fully formulated doubly quantized theory, 
a `state' can perhaps be thought of as a `superposition' of various `singly'
quantized states, each of which predicts definite probabilities. 
A `measurement' in a
`doubly' quantized theory can be expected to collapse the `doubly' quantized state to a
`singly' quantized one, selecting a particular probability distribution from all possible
ones. Every `measurement' will lead to a different probability distribution, so no
definite probability will be predicted.
On the other hand, the expectation value will be given by an average
over all the averages of the `singly' quantized probability distributions.

\subsection{New Phenomenology?}

In conclusion, let us offer 
some remarks on possible experimental observations of such super-quantum 
violations of Bell's inequalities in quantum gravity.

The usual experimental 
setup for testing the violation of Bell's inequalities in quantum mechanics
involves entangled photons \cite{exp}. In OSFT, photons are the lowest lying 
massless states, but there is a whole Regge trajectory associated 
with them. The obvious experimental suggestion is to look for 
effects from entangled Reggeized photons. Such experiments are of course 
impossible at present, given their Planckian nature.

A more feasible 
place to look for super-quantum correlations could be in 
cosmological data. It is believed that quantum fluctuations seed the 
large scale structure of the Universe, i.e. galaxies and clusters 
of galaxies that we observe.\cite{Bardeen:1983qw} The simplest models use Gaussian 
quantum correlations, though non-Gaussian correlations are envisioned as well 
and are constrained by data on the cosmic microwave 
background (CMB) from the Planck satellite.\cite{planck} 
While it is yet 
unclear how super-quantum correlations would affect the CMB data, 
we expect that they would leave ``stringy'' imprints on the 
large scale structure of the Universe and be observable at 
those scales.

Similarly, quantum gravitational imprints could be expected in the dark energy sector\cite{Chang:2011yt,Chang:2013lsa,Chang:2010ir,Chang:2011jj,Freidel:2013zga} as well
as in the dark matter \cite{darkm} and the Standard Model sectors.\cite{ufuk} 
If indeed quantum gravity demands a
new post-quantum framework for physics as we have argued in this talk, 
dramatic phenomenological consequences are to be expected at all scales of fundamental physics and cosmology.

\section*{Acknowledgments}

We would like to thank Nick Gray for helpful discussions. 
ZL and DM were supported in part by
the U.S. Department of Energy, grant DE-FG02-13ER41917, task A.
TT thanks the organizers of the Conference in Honor of the 90th birthday of Freeman Dyson
for the opportunity to present this talk.
TT is also grateful for the hospitality of the Kavli-IPMU during his sabbatical year from fall 2012 to summer 2013,
where he was supported by the World Premier International Research Center Initiative (WPI Initiative), MEXT, Japan.


\end{document}